# From Bits to Atoms:
# 3D Printing in the Context of Supply Chain Strategies

Henrik J. Nyman
Åbo Akademi University
Peter.Sarlin@abo.fi

Peter Sarlin
TUCS / Åbo Akademi University
Peter.Sarlin@abo.fi

## Abstract

*A lot of attention in supply chain management has been devoted to understanding customer requirements. What are customer priorities in terms of price and service level, and how can companies go about fulfilling these requirements in an optimal way? New manufacturing technology in the form of 3D printing is about to change some of the underlying assumptions for different supply chain set-ups. This paper explores opportunities and barriers of 3D printing technology, specifically in a supply chain context. We are proposing a set of principles that can act to bridge existing research on different supply chain strategies and 3D printing. With these principles, researchers and practitioners alike can better understand the opportunities and limitations of 3D printing in a supply chain management context.*

## 1. Introduction

Success or failure of companies can hinge on getting the right product, at the right time, and at the right price to customers. A magnitude of research and a lot of attention by practitioners has been devoted to this challenge. This has some answers regarding how to build supply chains (SCs) that can cater for demand for different products, timely delivery, or low cost, either as separate targets or simultaneously. But what if you could have infinite customization possibilities without a cost penalty? What if your production could take place in front of your customer at the push of a button? And what if neither finished goods inventory, nor a lengthy process for setting up production machinery would be needed?

Established research on SC strategies has identified different means for companies to cater for customer requirements. A SC strategy in our context is defined as the practical methods to achieve a given goal. This goal, driven by customer requirements, can be to provide for example low-cost products, customized products, high-quality products, or products that can be delivered quickly. In this paper, three different SC strategies are looked at: lean, agile, and a combination of both in so called "leagile" SCs. Recent advances in technology for additive manufacturing, also known as three-dimensional (3D) printing, has brought about changes to manufacturing and production at large, some of which are yet to come. Our intention is to assess the implications of these changes, as far as we can identify them today, on the underlying assumptions of SC strategies. As 3D printing is specifically a manufacturing technology, our focus is on manufacturing SCs (as opposed to service SCs). For the same reason, we concentrate on operative characteristics of different SC strategies. This includes aspects like timing of production, product properties, and positioning of inventory in the chain. For example, communication, trust, and information transparency can also be seen as crucial aspects of any well-functioning SC. Yet, a different manufacturing technology is unlikely to affect these. We view operative characteristics as a sub-set of the plethora of methods available to achieve a given goal for the SC. Operative characteristics are those methods that are directly tied to manufacturing technologies and processes, network design, and product architecture.

We build a conceptual model for 3D printing in a SC context. The nature of our conceptual model follows the definition in Meredith [22]. It implies that a broad range of research on the same topic is summarized and analyzed for common elements, combined and then extended. In our case, we analyze SC strategies as well as additive manufacturing and propose a model that explains how 3D printing can affect SC strategies. While there has been an early attempt to assess the impact of 3D printing on SC strategies [30], a thorough analysis of benefits and drawbacks of 3D printing and implications for various aspects of SC strategies is missing.

Sections 2 and 3 of the paper review research on SC strategies and additive manufacturing,

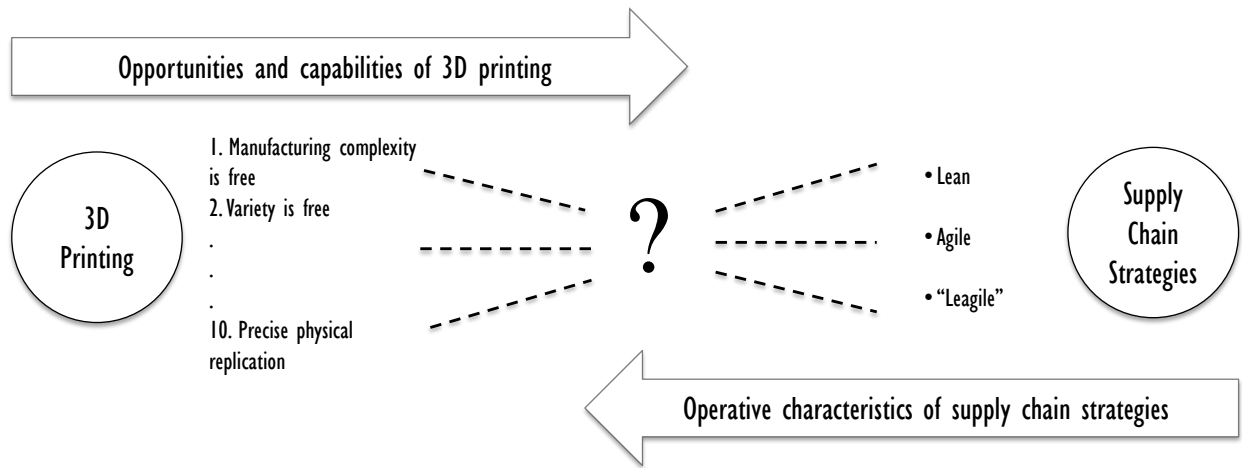

**Figure 1** The focus of the paper

respectively. For a comprehensive summary of 3D printing, we rely on the ten principles put forward by Lipson and Kurman [20], as is illustrated in Section 3 of the present paper. Section 3 also looks at the effects of 3D printing on the operative characteristics of any given SC strategy. In Section 4, we introduce the conceptual model of 3D printing in a SC context. As is illustrated in Figure 1, the purpose of the model is to connect the dots between operative characteristics of SC strategies, and capabilities of and opportunities in 3D printing. Section 5 contains a concluding discussion and directions for further research.

## 2. Supply chain strategies

The purpose of a manufacturing SC is to convert inputs in the form of raw material into parts or subassemblies, and eventually into a finished product which is delivered to the end customer. This typically involves several steps performed by multiple actors in the chain. SC management is commonly referred to as the co-ordination of transportation, information and money flows between involved parties (upstream to downstream and vice versa), so that the manufacturing and delivery of finished goods can be seen as one seamless process [12]. In addition to this transformative view of SCs, another key factor in catering for end-user requirements is to ensure that the variety of products reaching the market correspond to what consumers wish to purchase [13]. This can be seen on a product level (e.g., which car model does the consumer want?) or a product variety level (e.g., which color of a given car model does the consumer wish to purchase?). Failure to cater for variety can result in either lost sales opportunities, or goods sold at a discount with a small or non-existent profit margin. Understanding what to deliver and how to get it to the customer thus becomes important. This section has a twofold focus. First, Section 2.1 builds an understanding of different product types and their relation to SC strategies ("what to deliver"). Second, Sections 2.2, 2.3 and 2.4 look at SC strategies in more detail ("how to get your product to the customer").

## 2.1 Supply chain strategies for different product types

Goods sold to consumers can be categorized as either functional or innovative [13]. Innovative products imply products with a short-life cycle, high margin, and most importantly, unpredictable demand, while functional products are characterized by opposite attributes. This is not a black-and-white distinction, many products are offered as a "basic" variant that leans towards a more functional product while a "premium" variant of the same product can have characteristics typical for innovative products. The predictability of demand is a key factor in deciding on what to focus on in a SC for either functional or innovative products. In practice, predictable demand (for functional products) caters for opportunities in focusing on efficiency as the primary goal for the SC. A focus on efficiency can be gained, for example, through high capacity utilization, cost efficient transportation (e.g., sea freight as opposed to more expensive air freight), or lower inventories [13,6]. Innovative products with unpredictable demand on the other hand require a strategy that provides a higher degree of responsiveness in the SC. This implies, among other things, capacity as well as inventory buffers and more reactive (i.e., faster) transportation [13,6].

A similar analysis regarding the matching of product types to strategic goals for the SC is put forward by Mason-Jones et al. [21]. They discuss differences between so-called commodities, which are similar to functional products, and fashion goods that are similar to innovative products. They divide the specific goals for each product category into market qualifiers and market winners. Market qualifiers indicate metrics of high importance while market winner metrics are those where the SC must excel. Of note is that the market qualifiers for commodities and fashion goods are very similar: both should have an emphasis on quality and shorter lead-times. The distinguishing factors lie in service level (the market winner for fashion goods) and price (the market winner for commodities). This does not imply that price would be an insignificant factor for fashion goods (nor that service level is unimportant for commodities) – they are both in the market qualifier categories. As such, methods to fulfill particular goals are not mutually exclusive, but they have a different emphasis regarding the same goals.

A low price for the consumer requires low cost in the SC, or in other words, an efficient SC. The notion of service level is more multifaceted. We view it as a construct of two things: customization according to customer requirements and a timely delivery of the right goods. Customization can be translated to the number of products or variants on offer, or, in more extreme cases, the ability to offer fully tailored goods for one specific customer. A timely delivery implies a minimum lead-time that can be tolerated or a necessity for delivery within a given time frame. Some products display significant value loss unless available on the market before a given time. This can include goods that spoil (e.g., food) or seasonal items (e.g., Christmas cards). Ability for customization and timely delivery can jointly be viewed as a responsive SC. An SC strategy that emphasizes efficiency is frequently referred to as a lean SC, whereas a strategy that emphasizes responsiveness is frequently referred to as an agile SC [25,24]. The primary methods for achieving either efficiency or responsiveness are different, something we will look at in the next two sub-sections. Table 1 provides a summary of the discussion on different SC strategies, including their environment (related to the type of product in question), operative characteristics, consequences, and ultimate goal.

## 2.2 Lean supply chains

In a lean SC the primary focus is on ensuring a level schedule while eliminating all kinds of waste [25]. This results in a low cost for SC operations and a lower product price for the consumer. We primarily view a level schedule as a result of the nature of functional products. By their nature, they display less fluctuation of demand. However, through a focus on the elimination of waste, a level schedule can be promoted also for more complex products. In addition, the elimination of waste is at the core of cost effective operations at for example Toyota [10,19]. Waste is considered in broad terms, and Liker [19] defines eight different types of waste that should be eliminated in a lean setting. The last one is included for the sake of clarity, but given our focus on operative characteristics it will be excluded from subsequent analysis.

1. Overproduction – this involves production of goods for which there are no orders.
2. Waiting time – this can be caused by a number of reasons, such as stock-outs, processing delays, or capacity bottlenecks that cause idle production machinery.
3. Unnecessary transportation – moving finished goods or parts unnecessary long distances.
4. Over-processing or incorrect processing – any unneeded steps to produce goods, for example due to poor design. This can also involve producing goods at a higher degree of quality than necessary.
5. Excess inventory – this involves excess inventory of parts, sub-assemblies, or finished goods.
6. Unnecessary movement – this refers to the movement of personnel, everything should be easily accessible.
7. Defects – any type of rework or repair is considered waste. In other words, an appropriate product quality is needed. In addition, this involves scrap from low yield manufacturing processes.
8. Unused employee creativity – skills and ideas from employees should be used to the full.

Closely related to the topic of lean SCs are so-called green SCs. Society is increasingly putting pressure on companies to implement environmentally sustainable practices in their operations. Customer demand for environmentally friendly products and public pressure, combined with a high cost for material and energy are acting as incentives to ensure sustainability in operations [17]. Sustainable operations can be related to both manufacturing and transportation activities.

Research regarding the relationship between lean SC strategies and green operations show that there are many similarities in the two [16,23]. Specifically, the reduction of waste, in broad terms, acts as a mediator for environmentally friendly operations. This can be seen with, for example, less inventory and shorter transportation. In addition to this, the implementation of lean practices gives rise to a need for monitoring and controlling of operations. This can create opportunities in identifying measures to lessen environmental impact [23]. Thus, in many cases, lean is green.

### 2.3 Agile supply chains

Innovative products require a high service level and the agile SC is thus focusing on capturing market opportunities in a volatile market place [25]. The need for responsiveness can be linked to four constructs [28]: demand uncertainty, product variety, short lead times, and demand variability. Uncertainty regarding which product is desired at the market place and which particular variant of any given product is needed are prevalent for innovative products. In addition, the desire of customers to receive their product in a timely manner further creates pressure to react. Products with seasonal demand or short life cycles can also create a need for responsiveness even though demand in itself would be predictable. Rapid shifts in volumes can still occur under these circumstances. All in all, agile SCs operate in a volatile environment. Christopher [7] puts forward that the agile SC is characterized by its ability to respond to changes in demand, including both volume and variety. If we consider the operative factors involved in reaching this responsiveness, they can be built through a modular product architecture that provides possibilities for postponement and mass-customization, a reduction of product complexity that gives raise to greater manufacturing flexibility, and buffer inventory of either finished or semi-finished goods [28,7].

Postponement implies a delayed configuration of a given product, where the final assembly or customization takes place only once a customer order is received [7]. A product structure that allows for this is ideally based on modularization, where the final product is composed of different units that are assembled according to the preference of the buyer (e.g., a cover of a specific color is assembled to the final product based on customer preferences). This also allows for mass-customization, where a high number of different variants can be offered to customers while still retaining benefits of mass-production. For example, a single car model can be offered in many different variants to a customer.

Product complexity is on the one hand based on the number of variants or similar products on offer to customers and on the other hand on the level of non-standard parts used in a given product [7]. Fewer variants or products goes against the nature of agile SCs, but the key is in identifying products or variants that do not add value to the customer. Less can be more if it makes the ultimate choice for the consumer easier. Standardized products and fewer variants can be tied to manufacturing flexibility, which is defined as the length of changeover times for manufacturing machinery [28]. With fewer products in terms of finished goods or variants, there is a smaller need to move production from one product type to another. Also, products requiring less specialized production, such as those using the same parts or capacity as other similar products, allows for sharing of production capacity between multiple products. This reduces the need to change production from one part or product to another.

### 2.4 Leagile supply chains

Similarly as products cannot necessarily be categorized as only functional or innovative, lean and agile SC strategies have many similar goals. Leanness can be a part of agility and vice versa [8]. At the same time, what is considered waste in a lean setting (e.g., excess inventory or capacity), can be considered a necessity in an agile environment [21]. However, SCs are many times catered to simultaneously implement both lean and agile principles in so called "leagile" SCs [25,21,8]. A leagile SC makes use of operative characteristics of both lean and agile SCs to reach a balance between cost efficiency and responsiveness. Most commonly, this is achieved through the use of the order de-coupling point. The order de-coupling point separates the part of the SC where production is driven by forecast and the part where production takes place based on customer orders [8]. Lean principles are employed prior to the de-coupling point, while agile manufacturing is used for customer-order driven production after the de-coupling point. The use of a de-coupling point essentially coincides with the notion of postponement. Inventory is held in a generic form until a customer order is received.

There are however also other strategies that can be employed to achieve a leagile SC. In addition to a de-coupling point, separate processes for different products, or a separation of base demand and flexible demand for the same product can be employed [29]. Separate processes imply that different products (or different variants of the same product), are produced according to either lean or agile principles. The separation of base demand and flexible demand on the other hand implies that the predictable minimum demand for a given product is supplied with lean principles, whereas flexible agile principles exist to cover any upsurge of demand for the same product.

**Table 1** A summary of supply chain strategies, their starting point, operative characteristics and ultimate goal

|         | Environment | Operative characteristics | Consequences | Ultimate goal |
|---------|-------------|---------------------------|--------------|---------------|
| **Lean** | Level | Eliminate overproduction<br>Eliminate waiting time<br>Eliminate unnecessary transportation<br>Eliminate over or incorrect processing<br>Eliminate excess inventory<br>Eliminate unnecessary movement<br>Eliminate defects | Less waste | Cost efficiency |
| **Agile** | Volatile | Use of buffer inventories of goods<br>Modularization of products<br>Standardization of products<br>Fewer products or variants (that do not add value) | Availability of goods<br>Mass-customization and postponement<br>Manufacturing flexibility | Responsiveness |
| **Leagile** | Level and Volatile | Use of a de-coupling point<br>Separate processes for different products or variants<br>Separate base and surge demand for the same product or variant | Lean and agile combined | Balance between cost-efficiency and responsiveness |

## 3. Additive manufacturing and supply chain strategies

3D printing is a technology that uses an additive process for manufacturing three-dimensional objects from a digital model. This manufacturing technology uses a computerized design file to generate successive layers of the desired material. Rather than cutting away raw material or using molds, as is oftentimes the case in traditional manufacturing, it is thus additive rather than subtractive or formative. 3D printing has its origins in ink-jet printing technology developed in the late 1970s [11]. Similar to ink-jet printing, a printer head produces a layer. Unlike 2D printing, once this layer has solidified, a subsequent layer is built upon it to create the third dimension. While there are different specific technologies used in 3D printing, for example for different material (see e.g., [26] for an overview), the main operating principles are the same [27].

The core of 3D printing lies in the use of digital models stored as Computer Aided Design (CAD) files, which can either be created and designed with a modeling program for 3D objects or scanned into such a program with a 3D scanner. These CAD files are then split into digital cross-sections that are successively (additively) printed into 3D objects. When building a desired object, the cross-sections, or layers, may consist of liquid, powder, paper or sheet material, where printer resolution mainly relates to the thickness of each layer. Typically plastic is used as a raw material, but it is already now possible to form metal and other fabrics using 3D printers. All processes use either thermal energy or chemical reaction to bond the layers together [27]. This process is typically referred to as fusing, while the process in which a new layer is produced by the print-head is typically referred to as coating [26]. A crucial difference compared to traditional manufacturing methods lies in the absence of specific tooling or set-up of the production machinery. With a new CAD file, the 3D printer is instantly ready to produce a different item.

While the idea of additive manufacturing is not new, as the earliest publications of technologies for printing solid objects date back to the early 1980s (see e.g., [18]), large improvements have been made to the additive process during the past few years (see e.g., [3]). Additive manufacturing has been closely associated with rapid prototyping [1,5], but we are witnessing a move from prototyping to much more advanced applications of the technology in a broader manufacturing context [4]. This includes applications in aerospace and automotive parts production, tooling, biomedical parts, and artistic design, as reported by Petrovic et al. [26].

Many of the benefits with 3D printing relate to its additive, rather than subtractive, nature. This means that material waste is kept at a minimum, but also that complex shapes and forms can be produced without assembly [1,2,5,26]. In addition, without the need for custom tooling, the manufacturing process is fast to set up, and infinitely customizable. In practice, this implies on-demand, highly flexible production [1,2,5,26]. This can act to significantly reduce the need for buffer inventory of finished or semi-finished goods [2]. Another area of impact is in relation to sustainable operations. The technology allows for a high degree of recycling. If a low yield manufacturing process produces scrap, the material can simply be re-used in the manufacturing process [26]. Furthermore, the size of the printers themselves and the lack of tooling (e.g., molds for an injection molding machine) allow for a much more distributed manufacturing network. The only thing needed for production is a printer and a digitally distributed design file. While raw material still needs to be shipped to respective locations, there are typically better opportunities to source raw material locally compared to shipping of finished goods from bigger manufacturing facilities. Transportation and resulting carbon emissions can be reduced.

The technology, while developing at a fast pace, still however has a number of weaknesses prior to reaching its full potential. Hence, even though 3D printing is already in use and holds merit for a wide range of purposes, it is worth to note that it is no panacea from the viewpoint of production and manufacturing. From the literature, it is evident that the applications of 3D printing involve the manufacturing of short series of goods. This is related to the rather long time it takes to print an object [5], as well as to the nature of the process itself. Only a very small part of the manufacturing cost can be distributed across a large number of manufactured items. Essentially, the cost per item for producing 10 pieces is the same as for 10 000 pieces. The reported cost-effective quantities for 3D printed goods, compared to production of the same goods using plastic injection molding, range from 50 to 5000 units [2]. In addition to this, there are currently concerns regarding the lower precision of the manufacturing technique as well as regarding the limited choice of material and resulting shortcomings in the properties of the manufactured object [2,4]. Essentially, quality can be an issue.

That being said, Lipson and Kurman [20] outline ten principles that describe the possibilities this new technology has to offer. These principles encompass and summarize the opportunities of the technology in an appropriate way. Below, we start by briefly recapping the ten principles, and then provide a positioning of them and their opportunities in the context of different SC strategies. Table 2 provides a summary of the discussion. We will return to the limitations of the technology in Section 4.

*(1) Manufacturing complexity is free* This refers to the fact that there is no difference between simple and complex objects in the 3D printing world. In contrast to traditional manufacturing, a complicated shape or ornate is no more expensive than a simple block. From an SC strategy perspective, this takes customization to an entirely new level by allowing a low-cost agile strategy. In fact, this can be an example of "full" customization rather than mass-customization in the sense that a product can be fully customized without any restrictions. A modular product structure is no longer required. For example, ornaments or decorations to products can be produced in one-go with the same simplicity as a plain product. Thus, there is no need to have separate modules that are assembled differently according to customer preferences.

*(2) Variety is free* A 3D printer can in many cases be reconfigured to produce any different object in a minimal amount of time (and at a much lower cost) than traditional manufacturing machines. This implies not only rapid reconfiguration of machines, but also a lesser need for retraining of personnel. Similar to the point above, this allows for SC agility at a much lower cost than previously. We have concluded that a large number of varieties result in manufacturing inflexibility, due to the time penalty associated with production reconfiguration. This is no longer the case with 3D printing. Rapid reconfiguration of machinery also eliminates any possible waiting time (waste) associated with bottlenecks in the production.

*(3) No assembly required* Traditional manufacturing often relies on a complex set of SC partners delivering different parts to be assembled into a final product. As 3D printers create objects in layers, any interlocked parts can be produced "in one go", rather than relying on traditional assembly. This can eliminate entire steps in the manufacturing process, or even nodes in the SC. This greatly improves the possibilities to achieve a true lean SC strategy through less waiting time in the production process and less transportation required between parties in the chain. A more simple production process also removes any "unneeded steps" that result in over processing in production.

*(4) Zero lead time* 3D printing shortens the lead time between the initiation and execution of a

process. We view this as interrelated with, or even the result of, a number of other principles, such as portable manufacturing (7) and a simplified production process (3). This allows for very reactive manufacturing. Given that lead-times are crucial in both a lean and an agile setting, this will have an impact on both strategies. This affects overproduction, waiting time, and excess inventory as all of these are better managed with a very reactive production process. In addition, this combined with principles 3, 6, and 7 caters for a distributed production model. Delayed configuration can be taken to an entirely new level. As postponement is closely tied to the concept of the de-coupling point, principles 3, 6, and 7 also cater for leagility in that they provide extended flexibility in terms of positioning the de-coupling point. This extremely reactive production method can also allow for efficient separation of base demand and surge demand. A quick, reactive production method can be employed to take care of sudden demand increases.

*(5) Unlimited design space* Tools for traditional manufacturing are limited in terms of the shapes and forms they are able to produce. Each type of manufacturing has its own limitations, which may take the form of viable shapes or cost efficient production. With 3D printing, we remove some of these limitations, enabling entirely new designs. While having less direct impact on SC strategies, this can open up new possibilities for example in terms of customization.

*(6) Zero skill manufacturing* While handcrafted objects require a high degree of specific skills, manufacturing for mass-production requires less artisan skill, but still a high degree of specialization to operate the machinery used. 3D printing further reduces the skills needed by an operator, as a 3D printer is easier to operate compared to, for instance, an injection molding machine. In essence, traditional manufacturing often requires skills specific to the technology, whereas 3D printing is more versatile in that a wider variety of objects can be produced with the same technology (and thus skills). This allows for distributed production where, for example, final assembly takes place in the store where a customer purchases a product. In addition, the versatility of the production method can work to reduce over processing that, at its core, is about simplification of the production process.

*(7) Compact, portable manufacturing* 3D printers have larger manufacturing capacity per volume of production space compared to traditional manufacturing machinery. While traditional manufacturing machinery mostly create objects significantly smaller than themselves, 3D printers can not only fabricate objects as large as their print bed, but also objects that are larger than the printer (given that the printing apparatus moves freely). This aligns well with the lean philosophy where all kinds of waste should be removed, implying efficient use of space. Likewise, the compact nature of additive manufacturing can be thought of as providing additional flexibility in that production can be moved closer to the customer.

*(8) Less waste by-product* Traditional manufacturing technology, such as a lathe for shaping wood or metal, can create excessive material waste. Berman [2] reports that compared to traditional methods, 3D printing can reduce waste by 40% in metal machining. Lipson and Kurman [20] report even higher material waste levels in traditional manufacturing. This implies that much of the total raw material usage ends up on the factory floor with traditional manufacturing methods. From an SC strategy perspective, this goes to the core of the lean philosophy and can potentially have great impact on the total cost. Wasteful material usage can be seen as over processing, as well as producing unnecessary scrap (related to the 'defects' waste type).

**Table 2** 3D printing and areas of impact in supply chain strategies

|  | Lean | Agile | Leagile |
| --- | --- | --- | --- |
| **Manufacturing complexity is free** | - | (Mass-) Customization | - |
| **Variety is free** | Waiting time | Manufacturing flexibility | - |
| **No assembly required** | Waiting time, unnecessary transportation, incorrect processing | - | - |
| **Zero lead time** | Overproduction, waiting time, excess inventory | Postponement | Base and surge demand, decoupling point |
| **Unlimited design space** | - | (Mass-) Customization | - |
| **Zero skill manufacturing** | Incorrect processing | Postponement | Decoupling point |
| **Compact, portable manufacturing** | Unnecessary movement | Postponement | Decoupling point |
| **Less waste by-product** | Incorrect processing, defects | - | - |
| **Infinite shades of materials** | Waiting time, unnecessary transportation, incorrect processing | - | - |
| **Precise physical replication** | - | Manufacturing flexibility | - |

*(9) Infinite shades of materials* Mixing and blending of different raw material is difficult with traditional manufacturing techniques that mainly carve, cut or mold material into shape. While possibilities are currently limited also in the 3D printing space, Lipson and Kurman [20] estimate that blending different materials will be possible in the near future as 3D printing technology evolves. We see that this provides new possibilities to combine different steps in production, as well as potentially removing nodes in the SC (as certain sub-assemblies or parts are no longer sourced separately). The effects are similar to those of principle 3.

*(10) Precise physical replication* 3D printing will bridge the worlds of digital and physical objects. Design files can be endlessly replicated and distributed. In addition, while being at its infancy, 3D scanning will open up a new dimension to the replication of physical objects, particularly scanning, editing and duplicating in order to build precise replicas, or even improvements to the original physical objects. From an SC strategy perspective, design files can be quickly disseminated across the SC to rapidly initiate production, with no need for the configuration associated with traditional manufacturing machinery. Production can be reallocated at a moment's notice.

## 4. Connecting the dots: four principles for 3D printing and supply chain strategies

The above discussion on 3D printing in an SC context provided multiple insights to how this technology can affect SC strategies. With the ten principles of 3D printing, we get a more nuanced picture of the areas of impact. However, many of the principles overlap and support both strategies, in addition to there also being a lack of analysis of the shortcomings of 3D printing in SC strategies. The aim of the present section is to connect the dots between SC strategies and opportunities in, as well as limitations of, 3D printing. In this section, we attempt to crystallize the discussion in Section 2 and 3 into four general principles. As is shown in Figure 2, we describe 3D printing in the context of SC strategies in terms of the following four broad principles:

1. Green operations
2. A cost-efficient flexibility of the de-coupling point
3. A lack of economies of scale
4. How, where and who being redefined

These are not mutually exclusive, but subsequent principles build on the characteristics of the previous ones. For example, many of the aspects that contribute to green operations also result in flexibility regarding the de-coupling point. One aspect we have chosen to leave out is that of product quality, as it is highly context dependent and bound to change rapidly as 3D printing technology evolves.

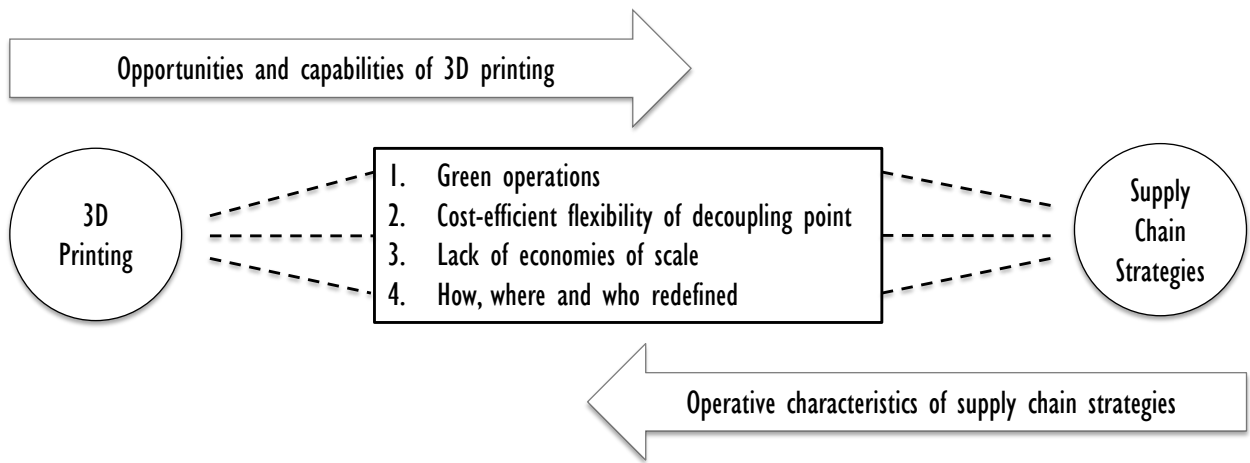

**Figure 2** Connecting the dots between 3D printing and supply chain strategies

Lean often supports sustainable operations. Given the high impact of 3D printing on lean characteristics in Table 2, we define the first new principle as green operations. Distributed, less wasteful manufacturing can have implications both on carbon emissions and material usage.

The second principle, cost-efficient flexibility of the de-coupling point, refers to the fact that a distributed manufacturing set-up can potentially bring production much closer to the customer. Without heavy costs in setting up capacity, related to for example skills and tooling, this is possible on a much bigger (or should we say smaller) scale than before.

The lack of economies of scale is of different nature in the sense that it, in most cases, can be perceived as negative. Despite the overwhelming support for lean SCs in Table 2, this principle actually implies that in many cases, 3D printing is *not* a viable option for a lean SC. This is due to the fact that traditional mass-produced goods have a cost advantage. The application of 3D printing technology must be carefully considered.

The last principle, how, where and who redefined, encompasses large implications for the SC. While it is unlikely that additive manufacturing will completely replace traditional manufacturing, it will complement means for traditional manufacturing. How things are produced is about to change, and this has implications that go beyond manufacturing. New skill-sets will be required related to for example CAD (where they previously might not have been needed). New ways to think about how we design and produce goods is required, and this will have an impact also on what we traditionally perceive as SC management expertise. Given the nature of 3D printing, distributed manufacturing can also eliminate entire nodes in the SC and bring about new business opportunities. Cottrill [9] reports on a third party logistics service provider (3PL) that shows great interest in 3D printing. This could entail 3PLs assuming an increasing role in manufacturing, in addition to their traditional role in distribution.

## 5. Conclusion

While the potential impact of 3D printing on SC strategies has lately been noticed in media, research in the field is still largely unexplored. With the help of the proposed conceptual model, we are able to point out opportunities in as well as limitations of 3D printing. As the technology evolves, there are bound to be changes to some of the underlying assumptions. Yet, the fundamental character of the technology will remain, such as the additive, rather than subtractive, process for creating an object.

The focus in this paper has specifically been on operative characteristics of SC strategies. As for example Cox [10] notes, there are many aspects of SC strategies that go beyond operative measures to more strategic areas like power, control and trust in the SC. The same can be said for 3D printing (or most likely for any new technology). Vinodh et al. [31] report on challenges related to managerial mindset and organizational culture in relation to the implementation of 3D printing for prototyping purposes. Furthermore, with possibilities for easy and rapid dissemination of design files, copyright issues related to 3D printing might arise [14]. This might become similar to what the music industry experienced with the onset of digital content. Subsequent research in many fields related to the topic in this paper is needed.

Holmström and Romme [15] discuss the nature of research in the field of operations management, in particular the misalignment of agendas in research and practice. They conclude that "practitioners do most of the activities that can be regarded as basic research, such as figuring out where and how novel technologies can be introduced to get operational benefits, and how novel technologies can be combined with existing operational practices in novel combinations". (p. 37). They, and us alike, see this as a challenge. In order for research to provide practical value, we need to explore new fields that are still to embody a large number of scholars and widespread research. This paper is an attempt at moving in that direction. While there are limitations in field-testing of the outlined principles in this paper, it serves as a starting point for further research. Work in the very near future aims at operatively evaluating the conceptual model presented in this paper.